\documentclass[11pt]{article}
\usepackage{graphicx}
\usepackage{amssymb,amsmath}

\begin{document}

\title{
\textbf{Power-law singularities and critical \\ exponents in $n$-vector models}
}

\author{J. Kaupu\v{z}s$^{1,2}$ \thanks{E--mail: \texttt{kaupuzs@latnet.lv}}
\\
$^1$ Institute of Mathematics and Computer Science, 
University of Latvia \\ 
29 Rai\c{n}a Blvd, LV--1459 Riga, Latvia \\ 
$^2$ Institute of Mathematical Sciences and Information Technologies, \\
University of Liepaja, 14 Liela Street, Liepaja LV--3401, Latvia}

\maketitle

\begin{abstract}
Power--law singularities and critical exponents in $n$--vector models are considered
from different theoretical points of view. It includes a theoretical
approach called the GFD (grouping of Feynman diagrams) theory, as well as the
perturbative renormalization group (RG) treatment.
A non--perturbative proof concerning corrections to scaling
in the two--point correlation function of the $\varphi^4$ model is provided, showing that predictions
of the GFD theory rather than those of the perturbative RG theory can be correct.
Critical exponents determined from highly accurate 
experimental data very close to the $\lambda$--transition point in liquid helium,
as well as the Goldstone mode singularities in $n$--vector spin models, evaluated from
Monte Carlo simulation results, are discussed with an aim to test the theoretical 
predictions. Our analysis shows that in both cases the data can be well interpreted within
the GFD theory.
\end{abstract}



\section{Introduction}
\label{sec:intro}

Critical phenomena in interacting many--particle systems
are associated with cooperative fluctuations of a large
number of microscopic degrees of freedom. Various physical quantities typically
exhibit power--law singularities in vicinity of the phase transition point. 
These are described by the critical exponents, which are exactly and rigorously known
for a class of exactly solved models~\cite{Onsager,McCoyWu,Baxter}. These are mainly the
two--dimensional lattice models. 
For three--dimensional systems, exact results are difficult to obtain,
and approximate methods are usually used -- see, e.~g., \cite{Amit,Ma,Justin,Kleinert,PV}
for a review  of the applied here standard perturbative renormalization group (RG) methods.
A general discussion of aggregation processes and critical phenomena in 
many--particle systems can be found, e.~g., in~\cite{Sornette,106.}. 
Recently, the role of quantum fluctuations in critical phenomena has been considered~\cite{Singh}.
For a general review, one has to mention that phase transitions described by
power laws and critical exponents can be observed in variety of systems, 
such as social, economical, biological systems, as well as vehicular traffic 
flow~\cite{nasch92,schad00,131.,148.}.

An alternative theoretical approach in determination of the critical exponents 
in the $\varphi^4$ model has been proposed in~\cite{K_Ann01}. We will call it the GFD theory, as it is based on certain grouping 
of Feynman diagrams. Moreover, a qualitative analysis is
performed here~\cite{K_Ann01} without cutting the perturbation series. More recently, this approach has been
generalized~\cite{K2010} to study the $\varphi^4$ model below the critical temperature,
where the so called Goldstone mode singularity (see,
e.~g.,~\cite{Sch,Law1,Law2,HL,Tu,SH78,ABDS99}) is observed. It refers to certain models, further called $n$--vector models, 
in which the order parameter is an $n$--component vector with $n>1$. This provides one more example of power law behavior, exhibited
by the transverse and longitudinal correlation functions in the ordered phase.
Moreover, according to the recent Monte Carlo (MC) simulation
results~\cite{KMR07,KMR08,KMR10}, it is very plausible that this behavior is described by nontrivial exponents,
as predicted in~\cite{K2010}. 
We will discuss this question in Sec.~\ref{sec:Goldstone}, including some new Monte Carlo simulation results.

\section{Critical exponents of the GFD theory}
\label{sec:crex}

Here we discuss the critical exponents of the GFD (grouping of Feynman diagrams)
theory~\cite{K_Ann01}, since these are important in our further tests and discussions. 
We consider a $\varphi^4$ model with the Hamiltonian
\begin{equation} \label{eq:H}
\frac{H}{k_B T}= \int \left( r_0 \varphi^2({\bf x}) + c (\nabla \varphi({\bf x}))^2 
+ u \varphi^4({\bf x}) \right) d {\bf x} \;,
\end{equation}
where the order parameter $\varphi({\bf x})$ is an
$n$--component vector with components $\varphi_i({\bf x})$, depending on the
coordinate ${\bf x}$, $T$ is the temperature, and $k_B$ is the Boltzmann constant.
As in~(\cite{K_Ann01}) and~(\cite{K2010}), it is assumed here that the order-parameter-field 
$\varphi_i({\bf x})$ does not contain the Fourier components $\varphi_i({\bf k})$
with $k> \Lambda$, i.~e., there exists the upper cut-off parameter $\Lambda$.
It is well known that the $\varphi^4$ model with certain dimensionality $n$ of the
order parameter belongs to the same universality class as the $n$--vector lattice spin model
with spins being $n$--component vectors of unit length (see Sec.~\ref{sec:Goldstone}). 
Therefore, the considered here predictions for the critical exponents refer also to these $n$--vector models.

It has been claimed in~\cite{K_Ann01} that possible values of the exact critical exponents for 
the $\varphi^4$ model in two ($d=2$) and three ($d=3$) dimensions are
\begin{equation} \label{eq:result}
\gamma = \frac{d+2j+4m}{d(1+m+j)-2j} \;; \hspace*{3ex}
\nu = \frac{2(1+m)+j}{d(1+m+j)-2j} \; .
\end{equation}
Here $\gamma$ is the susceptibility exponent,  $\nu$ is the correlation--length exponent,
$m$ may have a natural value starting with $1$, and $j$ is an integer equal or larger than $-m$.
Other critical exponents can be calculated from these ones, using the known scaling relations.
These values agree with the known exact solutions of the two--dimensional Ising model
($m=3$, $j=0$) and of the spherical model ($j/m \to \infty$). 
Based on the idea that $d$ might be considered as a continuous parameter in~(\ref{eq:result}) 
within $2 \le d \le 4$, a prediction has been made also
for the three--dimensional ($3D$) Ising model: $\gamma=5/4$ and $\nu=2/3$, 
corresponding to $m=3$ and $j=0$, as in the two--dimensional case.
This value of $\nu$ is consistent with the logarithmic singularity of specific
heat (according to $\alpha=2-d \nu =0$) proposed earlier by Tseskis~\cite{Tseskis}. 
The exponents $\gamma=5/4$ and $\nu=2/3$ have been later conjectured for the 
$3D$ Ising model by Zhang~\cite{Zhang}. 
The disagreement of these exponents with those of the perturbative RG method
can be understood based on a critical analysis~\cite{eprint}.
As explained in~\cite{K_Ann01}, the equations~(\ref{eq:result}) are meaningful
for positive integer $n$, and we can have $j=j(n)$ and $m=m(n)$ in the case where 
the order parameter is an $n$--component vector.

A relevant question is how~(\ref{eq:result}) can be related to specific models,
i.~e., one needs to find $j(n)$ and $m(n)$. It is well known that
the spherical model corresponds to the limit $n \to \infty$, so that we have $j(n)/m(n) \to \infty$
at $n \to \infty$ to obtain the known critical exponents of the spherical model in this limit. 
Consider now the critical exponent 
\begin{equation}
\eta = 2 - \frac{\gamma}{\nu} = \frac{4-d}{2(1+m)+j} \;,
\end{equation}
which describes the $\sim k^{-2+\eta}$ singularity (at $k \to 0$) of the 
Fourier--transformed two--point correlation function at the critical point.
Since $j(n)/m(n) \to \infty$, the asymptotic behavior of $\eta=\eta(n)$ for large $n$ is
\begin{equation}
\eta(n) \simeq \frac{4-d}{j(n)} \qquad \mbox{at} \quad n \to \infty \;.
\end{equation}

A general conjecture for $j(n)$ and $m(n)$ can be made assuming
that the critical exponents are given by analytic expressions valid for each positive
integer $n$. It results in an essential restriction on possible asymptotic form of $\eta(n)$.
In fact, the only reasonable possibility is $\eta(n) \propto n^{-s}$ at $n \to \infty$, 
where $s$ must be a positive integer, taking into account that $j(n)$ has an integer value
at any positive integer $n$. The latter means also that $j(n) \simeq a \, n^s$ holds at $n \to \infty$
with an integer coefficient $a$. An appropriate conjecture 
is $j(n)=\mathcal{P}_s(n)$ and $m(n)=\mathcal{Q}_r(n)$, where $\mathcal{P}_s(n)$
and $\mathcal{Q}_r(n)$ are polynomials of orders $s$ and $r<s$ with integer coefficients,
since $m(n)$ and $j(n)$ are integers at any integer $n \ge 1$ and
$j(n)/m(n) \to \infty$ holds at $n \to \infty$.
If we accept a physically reasonable idea that critical exponents and their derivatives with respect to 
$n$ behave smoothly and monotonously at $n \to \infty$, like some power of $n$, then this conjecture
is the only possible one. 
Hence, we have $n \, \eta(n) = \mathcal{F}(x)$, where $\mathcal{F}(x)$ is 
some analytic function of the argument $x=1/n$, which can be expanded
as $\mathcal{F}(x) = a_0 + a_1 x + a_2 x^2 + \ldots$ at small positive $x$. 
Here $a_0 \ne 0$ holds if $s=1$, whereas $s=2$ is a special case of $a_0=0$ and $a_1 \ne 0$.
Similarly, $s=3$ is a special case of $a_0=a_1=0$ and $a_2 \ne 0$, and so on.
Since we currently do not see any reason why the coefficient $a_0$ has to be zero,
we consider the choice $s=1$ as the most natural one, and we will further test only
this possibility. Thus, we have $j(n) = \mathcal{P}_1(n) = j(1) + a \, (n-1)$
and $m(n) = \mathcal{Q}_0(n) = m(1)$. Here $j(1)=0$ and $m(1)=3$ hold according to
the prediction for the Ising model. It yields $m \equiv 3$ and $j(n) = a \, (n-1)$ 
with some integer coefficient $a$. By definition, $j(n) \ge -m$ holds for $n \ge 1$ and, therefore,
$a \ge 1$. Only at $a=1$ we have a bijective relation between $j$ and $n$, i.~e., each nonnegative
integer $j$ corresponds to one positive integer $n$ and vice versa.
From this point of view, the choice $a=1$ seems to be the most natural one,
since in other cases a question arises why only each second (at $a=2$), each third (at $a=3$),
etc., integer value of $j$ has a meaning. We will further test just the possibility $a=1$,
i.~e., the conjecture 
\begin{equation}
m \equiv 3 \;; \hspace*{3ex}
j(n) = n -1 \;.
\label{eq:conj}
\end{equation}

Not only the critical exponents of leading singularities, but also
corrections to scaling are important in our tests.
According to~\cite{K_Ann01}, corrections to scaling can be represented by an expansion 
of correction factor (amplitude) in integer powers of $t^{2 \nu - \gamma}$ and 
$t^{2 \gamma - d \nu}$ at $t \to 0$, 
where $t$ is the reduced temperature. Since $(2 \gamma - d \nu)/(2 \nu - \gamma)$
is an integer number according to~(\ref{eq:result}), and $2 - \gamma/\nu = \eta$ holds according
to the known scaling relation, we obtain the expansion in powers of $t^{\eta \nu}$. 
Thus, the correction--to--scaling exponent is
\begin{equation}
\theta = \eta \nu = \omega \nu
\label{eq:correx}
\end{equation}
if the first expansion coefficient
is nonzero. Here $\omega = \theta / \nu$ is the correction--to--scaling exponent,
describing the $\sim k^{\omega}$ corrections to the critical correlation function.
It shows up also in the finite--size scaling analysis.
Allowing that some of the expansion coefficients are zero, we can have 
$\theta= \ell \eta \nu$, where $\ell= \ell(n)$ is a positive integer. A conjecture
$\ell(n) = 3 + n$ together with~(\ref{eq:conj}) has been tested earlier in~\cite{K05_he}, 
based on the experimental data near $\lambda$--transition point in liquid helium. 
This conjecture is based on the idea that $\theta$ for some quantities
tends to $1$ at $d \to 2$ in the Ising case of $n=1$, as well as at $d=3$ in the limit 
of the spherical model $n \to \infty$. 

A smooth crossover to the $2D$--Ising--behavior at $n=1$ and $d \to 2$,
or to the spherical--model--behavior at $n \to \infty$ can be expected in the 
$\varphi^4$ model. However, it is not necessary for the existence of such a crossover
that the correction--to--scaling exponent tends to the corresponding limit value. 
A simple possibility is that the expansion coefficients for the nontrivial
correction terms vanish in these limit cases. Moreover, even this is not the necessary
condition at $n \to \infty$. In fact, if we consider a correction factor of the
form $1 + \sum_{\ell=1}^M b_{\ell}(n) t^{\eta(n) \nu(n) \ell}$, where $M$ is an arbitrarily
large constant, then it is sufficient that 
$\lim_{n \to \infty} \sum_{\ell=1}^M \mid b_{\ell}(n) \mid < C$ holds for any $M \ge 1$,
where $C$ is a finite constant, which is independent of $M$.
It ensures that this correction factor tends to a finite constant at $n \to \infty$ for 
any given nonzero value of $t$, since $\eta(n) \to 0$ in this limit.
It means that the coefficients of the included here nontrivial correction terms
do not need vanish at $n \to \infty$.
As regards the $n=1$ case at $d=2$, it is further shown in Sec.~\ref{sec:nonp}
that nontrivial correction terms can be present in the $\varphi^4$ model and
absent in the $2D$ Ising model.

According to this discussion, Eq.~(\ref{eq:correx}) represents the most natural
conjecture for the correction--to--scaling exponents $\theta$ and $\omega$, since in this case we do not
need to assume that some of correction terms always vanish. This conjecture
is supported by the recent Monte Carlo simulation results~\cite{MC_Ising} for the 3D Ising
model on very large lattices with linear sizes up to $L=1536$.
Besides, according to the numerical transfer matrix calculations in~\cite{trfm}, a nontrivial correction to 
finite--size scaling  with the exponent $\omega=\eta$, probably, 
exists in the two--point correlation function even in the 2D Ising model.
The relation~(\ref{eq:correx}) is incompatible with the predictions of the
perturbative RG theory. However, an essential non--perturbative proof will be provided in
Sec.~\ref{sec:nonp}, showing that corrections to scaling of the GFD theory rather than those
of the perturbative RG theory can be correct.
In Sec.~\ref{sec:he}, we will test how well the conjectures~(\ref{eq:conj}) and~(\ref{eq:correx})
are consistent with the experimental data very close to the $\lambda$--transition point in 
liquid helium. It refers to the case $n=2$.

\section{Corrections--to--scaling theorem and a non--perturbative proof}
\label{sec:nonp}

Consider now the $\varphi^4$ model~(\ref{eq:H}) at $T \ge T_c$ (i.~e., $r_0 \le r_{0c}$), 
where $T_c$ is the critical temperature. We consider the case where $r_0$  is the only 
parameter which depends on $T$, and the dependence is linear.
In the following, we use the known thermodynamic relations for the free energy $F$, specific heat $C_V$,
internal energy $U$ and entropy $S$ at a fixed volume $V$:
\begin{eqnarray}
C_V &=& -T \left( \frac{\partial^2 F}{\partial T^2} \right)_V 
= - \frac{\partial}{\partial T} \left( T^2 \frac{\partial(F/T)}{\partial T}  \right)_V
=  \left( \frac{\partial U}{\partial T} \right)_V  \;, \label{CV} \\
U &=& F + TS = F -  T  \left( \frac{\partial F}{\partial T} \right)_V 
= - T^2 \left( \frac{\partial(F/T)}{\partial T} \right)_V \;,
\label{U}
\end{eqnarray}
as well as the well known relation $F = -k_B T \ln Z$, allowing to
determine the free energy $F$ from the microscopic model by calculating the
partition function $Z= \int \exp[-H/(k_BT)] \mathcal{D} \varphi$, where 
the symbol $\mathcal{D} \varphi$ indicates that
the integration takes place over all allowed configurations of $\varphi({\bf x})$. 
It yields
\begin{equation}
\frac{\partial}{\partial r_0} \left( \frac{F}{k_B T} \right) = - \frac{\partial \ln Z}{\partial r_0} 
= V \left\langle  \varphi^2({\bf x}) \right\rangle = n \sum\limits_{k<\Lambda} G({\bf k}) \;,
\label{der}
\end{equation}
where $G({\bf k}) = \langle \mid \varphi_i({\bf k}) \mid^2 \rangle$ (for any $i=1,2, \ldots, n$) 
is the Fourier--transformed (according to $\varphi_i({\bf x}) = V^{-1/2} \sum_{k<\Lambda}
\varphi_i({\bf k}) \, e^{i{\bf kx}}$) two--point correlation function. 

Let us denote by $U^{sing}$ and $C_V^{sing}$ the leading singular parts of 
$U/V$ and $C_V/V$ in the thermodynamic limit $V \to \infty$, 
represented in terms of the correlation length $\xi$  at $\xi \to \infty$, i.~e., at $T \to T_c$.
From~(\ref{U}) and~(\ref{der}), where $r_0$ is a linear function of $T$, we obtain
\begin{equation}
U^{sing}  \propto \left( \int_{k<\Lambda} G({\bf k}) d {\bf k} \right)^{sing}
= \left( \int_{k<\Lambda} [G({\bf k})- G^*({\bf k})] d {\bf k} \right)^{sing} \;,
\label{eq:Usi}
\end{equation}
where $(\cdot)^{sing}$ generally denotes the leading singular part of the corresponding
quantity in brackets, and $G^*({\bf k})$ is the correlation function at $T=T_c$.
The non-singular constant part $\int_{k<\Lambda} G^*({\bf k}) d {\bf k}$ 
is subtracted for convenience. 

Further on, we assume that $C_V$ has either the usually expected power--law singularity
$C_V^{sing} \propto \xi^{\alpha/\nu}$ or, more generally, a power--law singularity with logarithmic
correction of the form $C_V^{sing} \propto (\ln \xi)^{\lambda} \xi^{\alpha/\nu}$.
 Since $\xi \sim t^{-\nu}$ holds at $t \to 0$, Eq.~(\ref{CV}) then leads to
$C_V^{sing} \propto \xi^{1/\nu} U^{sing}$.
As it is well known, the critical long--wave fluctuations are responsible for the leading 
singularities near $T_c$. Hence, $U^{sing}$ cannot be altered by a short--wave contribution
to~(\ref{eq:Usi}), so that
\begin{equation}
U^{sing}  \propto \left( \int_{k<\Lambda'} [G({\bf k})- G^*({\bf k})] d {\bf k} \right)^{sing}
\label{eq:Usin}
\end{equation}
holds for any finite $\Lambda'<\Lambda$, i.~e., the leading singularity is independent of 
the upper integration limit, which is formally set to $\Lambda'$. Note, however, that 
$G({\bf k})$ and $G^*({\bf k})$ always correspond to the true upper cut-off $\Lambda$.
Summarizing these relations, we obtain
\begin{equation}
C_V^{sing} \propto \xi^{1/\nu} \left( \int_{k<\Lambda'} [G({\bf k})- G^*({\bf k})] d {\bf k} \right)^{sing} \;.
\label{eq:CVsing}
\end{equation}

Since only the small--$k$ contribution is relevant, it might be well justified to use 
the scaling hypothesis for $G({\bf k})$ and $G^*({\bf k})$, which is valid for 
small $k$ and large $\xi$. Namely, we have
\begin{equation}
G({\bf k}) = \sum\limits_{\ell \ge 0} \xi^{(\gamma - \theta_{\ell})/\nu} g_{\ell}(k \xi) \;,
\label{eq:sc1}
\end{equation}
where $g_{\ell}(k \xi)$ are scaling functions, $\theta_0=0$ holds and the term with $\ell=0$ 
describes the leading singularity, whereas the terms with $\ell \ge 1$ represent other singular contributions with
correction exponents $\theta_{\ell}>0$. Here we do not include possible analytic correction terms,
since they cannot give the leading singularities of $C_V$ and $U$.
The critical correlation function
\begin{equation}
G^*({\bf k}) = \sum\limits_{\ell \ge 0} b_{\ell} k^{(-\gamma + \theta_{\ell})/\nu}
\label{eq:sc2}
\end{equation}
is obtained at $\xi \to \infty$, so that we have
\begin{equation}
g_{\ell}(z) \simeq b_{\ell} z^{(-\gamma + \theta_{\ell})/\nu} \qquad \mbox{at} \quad z \to \infty \;.
\end{equation}
For complete formal correctness, one should note that the use of~(\ref{eq:sc1}) and~(\ref{eq:sc2})
is justified if it yields a $\Lambda'$--independent $C_V^{sing}$, since in this case
the calculated $C_V^{sing}$ is not modified by the short--wave (not--small $k$) contribution,
which is evaluated only approximately.

Now we are ready to formulate the main result of this section as the following theorem.
\vspace*{1ex}

\textbf{Theorem.} \hspace{0ex} 
\textit{
If  the leading singular part of specific heat $C_V^{sing}$
in the actually considered $\varphi^4$ model 
has the form  $C_V^{sing} \propto (\ln \xi)^{\lambda} \xi^{\alpha/\nu}$ (with $\lambda=0$
corresponding to the usual power--law singularity), if this singularity
is provided by the $\Lambda'$--independent small--$k$ contribution
to~(\ref{eq:CVsing}) with the scaling hypothesis (Eqs.~(\ref{eq:sc1}) and~(\ref{eq:sc2})) being valid
for $G({\bf k})$ and $G^*({\bf k})$, and if $\gamma + 1 -\alpha - d \nu >0$ holds,
then the two--point correlation function 
contains a correction--to--scaling term with certain exponent 
\begin{equation}
\theta_{\ell} = \gamma + 1 -\alpha - d \nu \;,
\label{eq:sakariba}
\end{equation}
corresponding to one of the terms with $\ell \ge 1$ in~(\ref{eq:sc1}).
} 

\vspace*{1ex}

 \textit{Proof.} 
Inserting~(\ref{eq:sc1}) and~(\ref{eq:sc2}) into~(\ref{eq:CVsing}) and changing the integration
variable to $y=k \xi$, we obtain
\begin{equation}
C_V^{sing} \propto \left( \sum\limits_{\ell \ge 0} 
\xi^{-d+(1+\gamma-\theta_{\ell})/\nu} F_{\ell}(\Lambda' \xi) \right)^{sing} \;,
\label{eq:CVs}
\end{equation}
where
\begin{equation}
F_{\ell}(z) = \int\limits_0^z y^{d-1} \widetilde{g}_{\ell}(y) dy \qquad \mbox{with} \quad 
 \widetilde{g}_{\ell}(y) = g_{\ell}(y) - b_{\ell} y^{(-\gamma + \theta_{\ell})/\nu} \;.
\label{eq:integ}
\end{equation}
Although we are interested only in the leading singularity of specific heat,
the correction terms have to be retained in~(\ref{eq:CVs}) at this step, since
some contributions can vanish after the integration and, therefore, a term with $\ell >0$ can be important.
Let us first assume that the leading singular part of $C_V/V$ is provided
by a single term with certain $\ell$. In this case,
$C_V^{sing}$ has the required form $C_V^{sing} \propto (\ln \xi)^{\lambda} \xi^{\alpha/\nu}$
only if $F_{\ell}(\Lambda'\xi) \sim  [\ln (\Lambda' \xi)]^{\lambda} (\Lambda' \xi)^{\mu}$
holds at $\xi \to \infty$ with some exponent $\mu$. 
Note that $\lambda=\mu=0$ is always true if the integral in~(\ref{eq:integ}) 
is convergent at $z=\infty$. Such $\widetilde{g}_{\ell}(y)$, which
ensures this property, certainly exists. It can be, e.~g., any function which decays as
$y^{-\sigma}$ at $y \to \infty$ with $\sigma>d$. Obviously, the leading term in 
$[\ln (\Lambda' \xi)]^{\lambda} (\Lambda' \xi)^{\mu} = 
[\ln \Lambda' + \ln \xi]^{\lambda} (\Lambda' \xi)^{\mu}$ is 
$\Lambda'$--independent only if $\mu=0$ holds, in which case we obtain
$C_V^{sing} \propto (\ln \xi)^{\lambda} \xi^{-d+(1+\gamma-\theta_{\ell})/\nu}$.
It is consistent with the required form at the condition~(\ref{eq:sakariba}). 
Such $\widetilde{g}_{\ell}(y)$, which gives $\lambda \ne 0$ and $\mu =0$ also exists:
it can be any function, decaying as $(\ln y)^{\lambda-1} y^{-d}$ at $y \to \infty$.
The possibility $\ell=0$ is excluded, since~(\ref{eq:sakariba}) and 
$\gamma + 1 -\alpha - d \nu >0$ cannot be simultaneously satisfied at $\ell=0$, 
as  $\theta_0=0$ holds by definition.
Thus, the leading asymptotic
term of the correlation function gives vanishing contribution 
to $C_V^{sing}$. It means that either
$\lim_{z \to \infty} F_0(z) =0$ holds, or the term with $\ell=0$ gives an analytic
(e.~g., constant or $\sim t$) contribution.  

Let us now consider a possibility
that two different terms with indices $\ell$ and $\ell'$ are equally important,
i.~e., proportional to each other at $\xi \to \infty$, and 
$C_V^{sing} \propto (\ln \xi)^{\lambda} \xi^{\alpha/\nu}$ holds. 
It is possible only if $F_{\ell}(\Lambda'\xi) \sim  [\ln (\Lambda' \xi)]^{\lambda} (\Lambda' \xi)^{\mu}$ and
$F_{\ell'}(\Lambda'\xi) \sim  [\ln (\Lambda' \xi)]^{\lambda} (\Lambda' \xi)^{\mu'}$
hold at $\xi \to \infty$ with $\mu - \theta_{\ell}/\nu = \mu' - \theta_{\ell'}/\nu$,
implying that $\mu \ne \mu'$ (since $\theta_{\ell} \ne \theta_{\ell'}$ holds by definition).
However, the obtained result for $C_V^{sing}$ is $\Lambda'$--dependent,
as it contains a factor of the form ${\Lambda'}^{\mu} + B {\Lambda'}^{\mu'}$ with $B \ne 0$.
Similarly, the result is $\Lambda'$--dependent if any larger number of terms are equally important. 
Consequently, only one term with certain $\ell \ge 1$ contributes to $C_V^{sing}$
at the conditions of the theorem. Hence, (\ref{eq:sc1}) contains the corresponding 
correction--to--scaling term with $\theta_{\ell}$ given by~(\ref{eq:sakariba}). 
$\Box$

The condition $\gamma + 1 -\alpha - d \nu >0$ is very meaningful, as it reduces to $\gamma >1$
according to the well known hyper-scaling hypothesis
\begin{equation}
\alpha + d \nu = 2 \;,
\label{hyper}
\end{equation}
and $\gamma >1$ really (or almost surely) holds for the $\varphi^4$ model within $2 < d < 4$.
Hence, our theorem has some important consequences listed below.
\begin{enumerate}
\item
If the hyper-scaling hypothesis~(\ref{hyper}) 
holds, then the statement~(\ref{eq:sakariba}) reduces to
\begin{equation}
\theta_{\ell} = \gamma -1 \;.
\end{equation}
\item
Since $\theta_{\ell}$ is one of the correction exponents, we have
$\theta \le \theta_{\ell} = \gamma + 1 - \alpha - d \nu$ for the leading correction--to--scaling
exponent $\theta$. It reduces to $\theta \le \gamma -1$ according to the hyper-scaling hypothesis,
this statement being valid if $\gamma>1$.
\item
The actual consideration allows a possibility that 
$C_V^{sing} \propto (\ln \xi)^{\lambda} \xi^{\alpha/\nu}$ holds
with an arbitrary value of $\lambda$. However, if the decay of $\widetilde{g}_{\ell}(y)$
at $y \to \infty$ is power--like, i.~e., $y^{-\sigma}$, then we have either $\lambda=0$
(at $\sigma>d$) or $\lambda=1$ (at $\sigma=d$).
\item
Since the $\varphi^4$ model belongs to the Ising university class at $n=1$,
we have $\theta_{\ell} = \gamma -1 = 3/4$ in two dimensions at $n=1$,
according to the known exact result $\gamma=7/4$ of the 2D Ising model.
Hence, a non-trivial correction to scaling with $\theta_{\ell} = 3/4$
exists (if the conditions of the theorem are satisfied)
in the correlation function of the $\varphi^4$ model at $n=1$ and $d=2$.
According to the known exact results (see, e.~g.~\cite{YP}), such a correction does not 
appear in the correlation function of the 2D Ising model on an infinite lattice.
Apparently, the 2D Ising model is a special case, where the non-trivial
corrections to scaling usually (but, probably, not always~\cite{trfm}) vanish.
\item
The statement that the term with $\ell=0$ gives vanishing contribution to $C_V^{sing}$
implies the existence of some cancellation mechanism in~(\ref{eq:CVsing}).
\end{enumerate}

The second consequence that $\theta \le \gamma -1$ hods at $\gamma>1$ 
is inconsistent with the predictions of the perturbative RG theory,
whereas the corrections to scaling of the GFD theory,  discussed
in Sec.~\ref{sec:crex}, completely agree with the proven here theorem.
There are no doubts that the conditions of this theorem
are very reasonable from the physical, as well as mathematical, point of view. 
Hence, the corrections to scaling of the GFD theory rather than those
of the perturbative RG theory can be correct.

\section{Best experimental evidences for the power--law singularities near the critical point} 
\label{sec:he}

There are a lot of different experimental evidences for power--law singularities
near phase transition points. However, it is not our aim to give an exhaustive review of
this topic, so that we will focus only on the best experimental evidences available.
These are basically the specific heat measurements in zero--gravity (space)
conditions~\cite{Lipa} very close to the $\lambda$--transition point in liquid helium.
These measurements are done with a high degree of accuracy much closer to
the critical point than in any other experiments or numerical simulations.
Due to this reason, it is widely accepted to consider them
as crucial tests of validity of the theoretical predictions for the critical exponents.
We will briefly discuss also the second--sound velocity measurements of superfluid
fraction~\cite{GA,GA1} near the $\lambda$--transition point in liquid helium.
It is widely accepted that the $\lambda$--transition is described by the critical
exponents of the $n$--vector model with $n=2$.

It has been found in~\cite{Lipa} that the experimental specific heat ($C_p$)
data for a wide range of reduced temperatures
$5 \cdot 10^{-10} \le t \le 10^{-2}$ below the $\lambda$--transition temperature
$T_{\lambda}$ can be well fit to appropriate ansatz of the perturbative RG theory,
providing an estimate $\alpha=-0.0127 \pm 0.0003$ of the specific heat critical exponent,
in a satisfactory agreement with the usual RG values.
An alternative ansatz, including logarithmic correction and critical exponents
$\alpha=-1/13$ and $\theta=5/13$ ($\theta=(n+3) \eta \nu$ within
the GFD theory) has been proposed in~\cite{K05_he}.
Here we will test a different possibility: $\alpha=2-d \nu =-1/13$ and $\theta=1/13$ given
by~(\ref{eq:result}), (\ref{eq:conj}) and (\ref{eq:correx}) at $n=2$. 

We start our analysis with a critical reconsideration of the fits, obtained by assuming the 
usual RG correction--to--scaling exponent $\theta \simeq 0.529$, i.~e., the same one
used in~\cite{Lipa}, where two slightly different ansatz
\begin{equation}
C_p = C_0 + At^{-\alpha} \left(1 + a_1 t^{\theta} + a_2 t^{2 \theta} \right)
\label{ansatz1}
\end{equation}
and
\begin{equation}
C_p = C_0 + At^{-\alpha} \left(1 + a_1 t^{\theta} \right) + a_2 t 
\label{ansatz2}
\end{equation}
have been considered (in somewhat different notations than here) 
with $t=1-T/T_{\lambda}$ being the reduced temperature at $T < T_{\lambda}$.
In our fits, we have used the raw data of~\cite{Lipa}, as well as
certain binning of these data, described in~\cite{K05_he}.
This binning procedure differs slightly fro that one used in~\cite{Lipa}.
In fact, the binned data correspond within the error bars
to $\bar{C_p}(\bar{t}) = (b t)^{-1} \int\limits_{t}^{(1+b)t} C_p(\tau) d \tau$ 
with $\bar{t}=t(1+b/2)$ and some constant $b$, defining the averaging
interval of the binning procedure. It is easy to verify that, if $C_p(t)$ is given
either by~(\ref{ansatz1}) or by~(\ref{ansatz2}), then $\bar{C_p}(t)$ also is described
by the corresponding ansatz with only slightly different coefficients $A$, $a_1$ and $a_2$.
Therefore, the raw data and also the binned data can be fit to determine the 
critical exponent $\alpha$. Since the variation of $C_p$ within one binning interval
is very small, the raw data and the binned data lie practically on the same curve, and
the fit results are consistent within the error bars.
We have used both data sets to verify the robustness of our fitting procedures
and related analyses.

The fit of raw data to~(\ref{ansatz1}) over the whole range of the reduced temperatures 
$4.75 \times 10^{-10} \le t \le 9.52 \times 10^{-3}$ with fixed $\theta = 0.529$ yields
$\alpha = -0.01263(20)$, $C_0=460.5(6.2)$, $A=-447.6(6.1)$, $a_1=-0.0156(13)$ and $a_2=0.3306(96)$,
where the standard errors are indicated in brackets 
(meaning $\alpha = -0.01263 \pm 0.00020$, $C_0=460.5 \pm 6.2$, etc.). These values are
well consistent with those reported in~\cite{Lipa}.

\begin{figure}
\begin{center}
\includegraphics[width=0.7\textwidth]{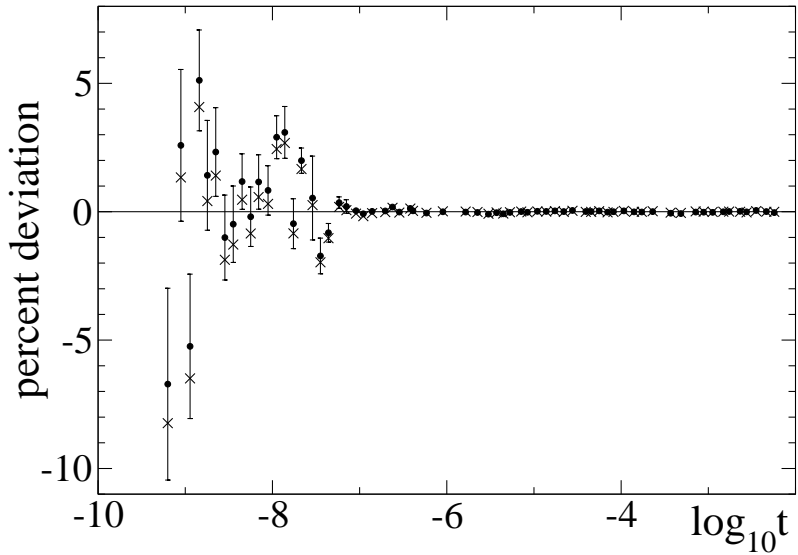}
\end{center}
\vspace*{-1ex}
\caption{The percent deviation of the measured (binned) $C_p$ data points from the fit
curve, obtained by fitting these data to~(\ref{ansatz1}) with fixed exponents 
$\alpha=-1/13$, $\theta=1/13$ (solid circles) and $\alpha=-0.01284$, $\theta=0.529$ (x).}
\label{percdev}
\end{figure}

We have performed certain test of validity of such a fit, based on the following idea:
this fit result for $\alpha$ should be consistent with the one obtained from a simpler ansatz
\begin{equation}
C_p = C_0 + At^{-\alpha} 
\label{simpleansatz}
\end{equation}
by fitting the data within $t \le t_{max}$ at so small values of $t_{max}$, 
at which the corrections to scaling $a_1 t^{\theta}$ and $a_2 t^{2 \theta}$ become negligible.
The effect of these correction terms is evaluated as $\alpha_1-\alpha_2$, where
$\alpha_1$ and $\alpha_2$ are obtained by fitting the $C_p$ data and the 
$\widetilde{C}_p = C_p - A \left( a_1 t^{\theta} + a_2 t^{2 \theta} \right)$
data, respectively, to the ansatz~(\ref{simpleansatz}). The latter data
are obtained by subtracting correction terms, evaluated from the overall fit to~(\ref{ansatz1}).
As soon as the difference between $\alpha_1$ and $\alpha_2$ becomes much smaller than
the error bars, a good agreement of 
$\alpha_1$ and $\alpha_2$ with the value $\alpha = -0.01263(20)$
of the overall fit is expected, if this fitting procedure with $\theta=0.529$ is valid.
The fit results for $\alpha_1$ and $\alpha_2$ depending on $t_{max}$ are represented in Tab.~\ref{tab1}.
\begin{table}
\caption{The estimates $\alpha_1$ and $\alpha_2$ of the critical exponent $\alpha$, 
obtained by fitting the $C_p$ data (for $\alpha_1$) and 
the $\widetilde{C}_p = C_p - A \left( a_1 t^{\theta} + a_2 t^{2 \theta} \right)$
data (for $\alpha_2$) to the ansatz~(\ref{simpleansatz}) within $t \le t_{max}$.}
\label{tab1}
\begin{center}
\begin{tabular}{|c|c|c|}
\hline
\rule[-2mm]{0mm}{7mm} 
$t_{max}$ & $\alpha_1$ & $\alpha_2$  \\
\hline
$3.18 \times 10^{-3}$ & -0.013950(58) & -0.012578(58) \\
$1.03 \times 10^{-3}$ & -0.012371(87) & -0.012724(87) \\
$3.70 \times 10^{-4}$ & -0.01194(17)  & -0.01270(17)  \\
$1.07 \times 10^{-4}$ & -0.01216(30)  & -0.01277(30)  \\
$2.95 \times 10^{-5}$ & -0.01135(53)  & -0.01179(53)  \\
$1.01 \times 10^{-5}$ & -0.00924(99)  & -0.00955(99)  \\
\hline
\end{tabular}
\end{center}
\end{table}
As we can see, the values of $\alpha_1$ and $\alpha_2$ very well agree within the 
error bars for the smallest $t_{max}$ value $1.01 \times 10^{-5}$. A good agreement
with the estimate $\alpha = -0.01263(20)$, however, is not observed at this $t$.
The disagreement cannot be reasonably explained by a possible inaccuracy in the 
$T_{\lambda}$ value. The discrepancy becomes smaller only by 
$0.0005$ if we shift $T_{\lambda}$ by $-0.5$~nK within the experimental
error bars. Moreover, the shift $T_{\lambda} \to T_{\lambda} - 0.5$~nK 
makes the deviations of the binned data points from the overall fit curve
quite remarkable for $10$ smallest $t$ values (see Fig.~\ref{percdev}). Namely, such a shift increases
the average deviation for these data points from $-1.10(70)$ percents to $-1.73(70)$ percents
and, therefore, is not well justified.
Hence, the actual fitting procedure with $\theta=0.529$ is doubtful.

In fact, if we assume that $\theta \simeq 0.529$ really holds,
then the fit to simple ansatz~(\ref{simpleansatz}) at very small $t$ values, such as $t \le 1.01 \times 10^{-5}$,
should be considered as a more reliable method than the overall fit within $t \le 9.52 \times 10^{-3}$
with two correction terms included. Indeed, it is expected that 
these corrections, as well as higher--order correction terms are negligible in the first case, 
whereas this is not surely true for the neglected higher--order corrections in~(\ref{ansatz1}) at
$t \sim 10^{-2}$ used there. From this point of view, $\alpha = -0.00924(99)$ is almost
the best estimate obtained by us, assuming $\theta \simeq 0.529$.
It does not well agree with the perturbative RG values, e.~g.,
$\alpha = -0.01294 \pm 0.0006$ reported in~\cite{Kleinert99}. 

The same test can be performed for the ansatz~(\ref{ansatz2}), yielding not better
results: we obtain $\alpha = -0.01320(21)$ from the overall fit to~(\ref{ansatz2}), this
estimate being not well consistent with $\alpha_1=-0.00924(99)$ and $\alpha_2= -0.00970(99)$ 
obtained from~(\ref{simpleansatz}) at  $t_{max} = 1.01 \times 10^{-5}$.
These tests have been performed also for the fits of the binned $C_p$ data, giving similar results.
For example, the overall fit to~(\ref{ansatz1}) in this case yields $\alpha=-0.01284(34)$, whereas
the fits to~(\ref{simpleansatz}) give us $\alpha_1 = -0.0081(13)$ and $\alpha_2=-0.0085(13)$
at $t_{max} = 1.09 \times 10^{-5}$. 
The bin--averaging has been performed with equal weights for all data points of one bin,
which is the reason for slightly larger statistical errors than for the raw--data fits.

A reasonable explanation of the failure in the above tests is provided by the non--perturbative
analysis in Sec.~\ref{sec:nonp}. Namely, the correction--to--scaling exponent
$\theta$ is likely to be not larger than $\gamma -1$. The latter quantity does not exceed $0.32$
according to both the perturbative RG estimate $\gamma = 1.3169 \pm 0.0020$ of~\cite{GJ98}
and our result in Sec.~\ref{sec:crex} $\gamma = 17/13 \simeq 1.3077$ for $n=2$.
Thus $\theta$ should be remarkably smaller than $0.529$, which means that the influence 
of the correction terms is still not negligible at $t$ about $10^{-5}$.

In the following, we test how well the data are described by the ansatz~(\ref{ansatz1})
with $\alpha=-1/13$ and $\theta=1/13$ proposed by~(\ref{eq:result}), (\ref{eq:conj})  
and (\ref{eq:correx}). Here we consider deviations from fit curves,
therefore the binned data are appropriate for the analysis, as they are less
noisy than the raw data.  We have found that the data can be well fit with fixed
$\alpha=-1/13$ and $\theta=1/13$ within $t \le 5.71 \times 10^{-3}$, as it is evident from
Fig.~\ref{percdev}, where the percent deviations from the fit curve are shown by solid circles.
The fit is less perfect for two largest $t$ values, which are omitted here.
However, since $\theta=1/13$ is quite small, it can be well explained by
an influence of higher--order correction terms. For comparison, the percent deviations
in the case of fixed exponents $\alpha=-0.01284$ and $\theta=0.529$ considered before
are shown by symbols ``x''. In fact, the percent deviations are practically the same 
in both cases for $t > 10^{-7}$.  The average deviation 
for 10 smallest $t$ values is $-0.10(70)$ at  $\alpha=-1/13$ and $\theta=1/13$,
i.~e., remarkably smaller in magnitude than $-1.10(70)$, obtained at $\alpha=-0.01284$ and $\theta=0.529$.
Thus, it might be true that the $t \to 0$ asymptotic is better described by the GFD exponents 
than by the perturbative RG exponents.

We have considered a series of fits of the binned $C_p$ data within $t \le t_{max}$ with fixed 
correction--to--scaling exponent $\theta=1/13$ and $\alpha$ as a fit parameter.
Since $\theta=-\alpha$ holds at $n=2$ according to~(\ref{eq:result}), 
(\ref{eq:conj}) and (\ref{eq:correx}), we have performed also fits with $\theta=-\alpha$.
These fits give similar results as those
with $\theta=1/13$, but the statistical errors are remarkably smaller for
very small $t_{max}$ values. Therefore, they provide a more precise test of the GFD theory. 
We have found that such fits are sufficiently (acceptably) stable within
$1.74 \times 10^{-5} \le t_{max} \le 6.99 \times 10^{-5}$,
as shown in Tab.~\ref{tab2}.
\begin{table}
\caption{The parameters of the binned--data fits 
to~(\ref{ansatz1}) within $t \le t_{max}$ with $\theta=-\alpha$.}
\label{tab2}
\begin{center}
\begin{tabular}{|c|c|c|c|c|c|}
\hline
\rule[-2mm]{0mm}{7mm} 
$10^5 t_{max}$ & $\alpha$ & $C_0$ & $A$ & $a_1$ & $a_2$  \\
\hline
$6.99$ & -0.069(17) & 201(31) & -474(29)  & -1.19(22) & 0.66(23) \\
$5.60$ & -0.069(18) & 201(33) & -474(30)  & -1.19(23) & 0.66(25) \\
$3.93$ & -0.080(20) & 185(26) & -473(19)  & -1.36(33) & 0.86(40) \\
$2.76$ & -0.069(30) & 200(51) & -461(30)  & -1.15(46) & 0.62(46) \\
$1.74$ & -0.091(27) & 175(27) & -494(44)  & -1.60(57) & 1.18(80) \\
\hline
\end{tabular}
\end{center}
\end{table}
The fitted $\alpha$ values in Tab.~\ref{tab2} are not very accurate. Nevertheless, they
perfectly agree within the error bars with the expected theoretical (GFD)
asymptotic value $-1/13 \simeq -0.07692$. The fits at three largest
$t_{max}$ values in Tab.~\ref{tab2} are sufficiently stable 
with respect to the temperature shift $T_{\lambda} \to T_{\lambda} \pm 0.5$~nK,
and the shifted $\alpha$ estimates agree within the error bars with $-1/13$.
We have fitted also the raw data to~(\ref{ansatz1}) with $\theta=-\alpha$.
Although the results are somewhat less stable in this case, the agreement
for very small $t_{max}$ values is observed, e.~g.,
$\alpha=-0.054(27)$ at $t_{max}=3.93 \times 10^{-5}$, 
$\alpha=-0.071(24)$ at $t_{max}=2.20 \times 10^{-5}$ and
$\alpha=-0.071(33)$ at $t_{max}=1.59 \times 10^{-5}$.

In fact, at so small 
$\theta$ as $1/13$, we practically cannot find such a $t_{max}$ value at which the
corrections to scaling are negligible. Even including two correction
terms, the ansatz is expected to be really accurate only at very small $t$ values.
Hence, the agreement with the theoretical prediction $\alpha = -1/13$ observed here at rather small $t$ values 
is an evidence in favor of the result (\ref{eq:result}) of the GFD theory  
and our conjectures (\ref{eq:conj}) and (\ref{eq:correx}).
The latter one might be more appropriate than 
$\theta = (n+3) \eta \nu$  used in~\cite{K05_he},
since now it is not necessary to assume the existence of a logarithmic correction
for a good fit of the data.

Consider now the superfluid--fraction data
of~\cite{GA,GA1}. It has been shown in~\cite{K05_he} that the effective 
exponent $\nu_{\mathrm{eff}}$, extracted from these data and plotted versus $t^{5/13}$, 
increases above the usual RG value $\nu \approx 0.67$ and apparently 
converges to a value near $9/13$ (the GFD theoretical value at $n=2$), if fitted to a parabola.
It is an evidence in favor of the conjecture $\theta = (n+3) \eta \nu$, the exponents 
$\eta$ and $\nu$ being consistent with~(\ref{eq:result}) and (\ref{eq:conj}).
This behavior can be explained assuming~(\ref{eq:correx}), as well.
Namely, first few coefficients in expansion of the 
superfluid fraction in powers of $t^{\eta \nu}$
can be very small, in which case the scaling of the data in a wide range of $t$ values is described by an effective 
correction--to--scaling exponent  $\theta_{\mathrm{eff}} \approx \ell \eta \nu$ with $\ell>1$.
Taking into account the experimental errors and uncertainty in the critical temperature
(it depends on pressure, which is not strictly constant in non--zero gravity conditions),
$\ell$ can be, e.~g., $5$, $4$, or even $3$ within the error bars.
Owing to the recent MC analysis of the
3D Ising model ($n=1$) on very large lattices~\cite{MC_Ising},
the conjecture~(\ref{eq:correx}) is more plausible than $\theta = (n+3) \eta \nu$.

In summary, it is possible to give a reasonable interpretation of the actually discussed
experimental and numerical (MC) data, assuming that 
(\ref{eq:result}), (\ref{eq:conj})  and (\ref{eq:correx}) hold.

\section{Goldstone mode singularities in the $O(n)$ models}
\label{sec:Goldstone}

Here we consider a class of $n$--vector spin models, where the spin is an
$n$--component unit vector with $n \ge 2$. These are also called $O(n)$ models due to the $O(n)$
global rotational symmetry exhibited by such $n$--vector  model in absence of the 
external field. The Hamiltonian $\mathcal{H}$ in this case reads
\begin{equation}
\frac{\mathcal{H}}{k_B T}=-\beta \left( \sum\limits_{\langle i j \rangle}
{\bf s}_i {\bf s}_j + \sum_i {\bf h s_i} \right) \;,
\end{equation} 
where ${\bf s}_i$ is the spin variable  of the $i$th lattice site, $\beta$ is the 
coupling constant, and ${\bf h}$ is the external field. The summation takes place
over all pairs $\langle i j \rangle$ of the nearest neighbors in the lattice.
Each spin has the longitudinal component $s_{\parallel}$, which is its projection on the
external field, and $n-1$ transverse components, which are perpendicular to the field.
We are interested in the magnetization per spin $M= \langle s_{\parallel} \rangle$, 
as well as in longitudinal and transverse correlation functions.
In the coordinate representation, the longitudinal ($\tilde G_{\parallel}({\bf x})$)
and the transverse ($\tilde G_{\perp}({\bf x})$) correlation functions are defined by
\begin{eqnarray}
\tilde G_{\parallel}({\bf x}_2-{\bf x}_1) &=& \langle s_{\parallel}({\bf x}_1) 
s_{\parallel}({\bf x}_2) \rangle -M^2  \label{eq:Gl} \\
\tilde G_{\perp}({\bf x}_2-{\bf x}_1) &=& \langle s_{\perp}({\bf x}_1) s_{\perp}({\bf x}_2) \rangle \;,
\label{eq:Gti}
\end{eqnarray}
where $s_{\perp}$ is any one of the transverse components.
Due to the symmetry of the model, the correlation functions depend only on the coordinate difference 
${\bf x}_2-{\bf x}_1$. 
The Fourier--transformed longitudinal and transverse correlation functions are 
\begin{eqnarray}
G_{\parallel}({\bf k}) &=& N^{-1} \sum\limits_{\bf x} \tilde G_{\parallel}({\bf x}) e^{-i{\bf kx}} 
\label{eq:GFpar} \\
G_{\perp}({\bf k}) &=& N^{-1} \sum\limits_{\bf x} \tilde G_{\perp}({\bf x}) e^{-i{\bf kx}} 
\label{eq:GFperp} \;.
\end{eqnarray}

Consider now the behavior of an $O(n)$ model below the critical temperature, 
i.~e., at $\beta>\beta_c$, 
in the thermodynamic limit $L \to \infty$. In this case, the magnetization $M(h)$ and the 
correlation functions exhibit Goldstone mode power--law singularities:
\begin{eqnarray}
&&M(h) - M(+0) \propto h^{\rho} \quad \mbox{at} \quad h \to 0 \;, \\ 
&&G_{\perp}({\bf k}) \propto k^{-\lambda_{\perp}} \hspace{5ex} \mbox{at} \quad h=+0 \quad \mbox{and} \quad k \to 0 \;,\\
&&G_{\parallel}({\bf k}) \propto k^{-\lambda_{\parallel}} \hspace{6ex} \mbox{at} \quad h=+0 \quad \mbox{and} \quad k \to 0 \;.
\end{eqnarray}
According to the standard theory~\cite{Law2,HL,Tu,SH78,ABDS99}, $\lambda_{\perp} = 2$ and $\lambda_{\parallel}=4-d$ 
hold for $2<d<4$, and $\rho = 1/2$ is true in three dimensions.
More nontrivial universal values are expected according to~\cite{K2010}, such that
\begin{eqnarray}
&&d/2 < \lambda_{\perp} < 2 \;, \label{eq:pred1} \\
&&\lambda_{\parallel} = 2 \lambda_{\perp} - d \;, \label{eq:pred2} \\
&&\rho = (d/\lambda_{\perp})-1  \label{eq:pred3}
\end{eqnarray} 
hold for $2<d<4$. These relations have been obtained in~\cite{K2010} by analyzing self-consistent
diagram equations for the correlation functions without cutting the perturbation series. 
Apart from the mathematical analysis, reasonable physical arguments also have been provided
there to show that $\lambda_{\perp}=2$ could not be the correct result for the $XY$ model
within $2<d<4$. 

The relations~(\ref{eq:pred1}) and~(\ref{eq:pred2}) are confirmed by MC
simulation results for the longitudinal and transverse correlation functions in the 
3D $O(4)$ model~\cite{KMR10}, where an estimate $\lambda_{\perp}=1.955 \pm 0.020$ has been found.
It has been stated~\cite{KMR10} that the behavior of the longitudinal correlation function
is well consistent with $\lambda_{\parallel}$ about $0.9$, in agreement with~(\ref{eq:pred2}) at
$\lambda_{\perp}$ about $1.95$, but not with the standard--theoretical prediction $\lambda_{\parallel}=1$.
According to~(\ref{eq:pred3}), we have $1/2 < \rho <1$ in three dimensions. A reasonable numerical evidence
for this relation  has been obtained in~\cite{KMR08} from the susceptibility
data of the 3D $XY$ model, providing the MC estimate $\rho = 0.555(17)$ for the $n=2$ case.
It corresponds to $\lambda_{\perp} = 1.929(21)$ according to~(\ref{eq:pred3}).

\begin{figure}
\begin{center}
\includegraphics[width=0.7\textwidth]{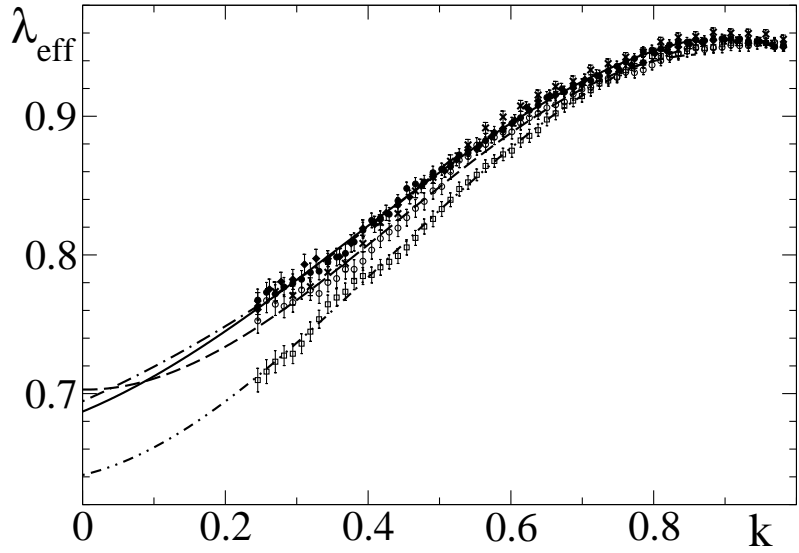}
\end{center}
\vspace*{-1ex}
\caption{The effective longitudinal exponent 
$\lambda_{\mathrm{eff}}$ depending on the wave vector magnitude $k$ at
$L=512$ and $h=0.000875$ (empty squares), $L=512$ and $h=0.0004375$ (empty circles),
$L=512$ and $h=0.00021875$ (solid circles), $L=384$ and $h=0.00021875$ (solid diamonds),
as well as at $L=256$ and $h=0.00021875$ (x). 
The cubic fits of the $L=512$ data within $k \in [k_{20},k_{70}]$
at $h=0.000875$, $h=0.0004375$ and $h=0.00021875$ are shown by
dot--dot--dashed line, dashed line and solid line, respectively.
The dot--dashed line shows the quadratic
fit of the latter data set within $k \in [k_{20},k_{43}]$.}
\label{effexpo}
\end{figure}

The longitudinal and transverse correlation functions for the 3D $XY$ model have not been 
simulated in the mentioned here papers. Here we fill this gap, providing a new MC evidence 
that $\lambda_{\parallel}<1$ holds. The simulations for several linear lattice sizes $L \le 512$
at $h=0.000875$, $h=0.0004375$ and $h=0.00021875$ have been performed at $\beta=0.55$ (the critical
coupling being $\beta_c \approx 0.4542$~\cite{SM}), and the correlation functions 
(in $\langle 100 \rangle$ direction) have been evaluated, following the method described in~\cite{KMR10}. 
As in~\cite{KMR10}, we evaluate the 
effective longitudinal exponents $\lambda_{\mathrm{eff}}(k)$ from the linear log--log fits within $[k,2k]$.  
The effective exponents depend also on $h$ and $L$, therefore plots of $\lambda_{\mathrm{eff}}(k)$ for
different values of these parameters are compared in Fig.~\ref{effexpo} to judge about the
asymptotic exponent $\lambda_{\parallel}$, corresponding to the limit 
$\lim\limits_{k \to 0} \lim\limits_{h \to 0} \lim\limits_{L \to \infty}$.
The finite--size effects increase with decreasing of $h$. We observe that the 
plots of the effective exponent for $L=512$ and $L=384$ lie almost on top
of each other at the smallest $h$ value, and the plot for $L=256$ also shows
only small deviations. Thus, the thermodynamic limit is practically reached
at $L=512$ for the actual values of the field $h$. Furthermore, the finite--$h$
effects are quite small in the considered range of wave vector magnitudes $k \ge k_{20}$,
where $k_{\ell} = 2 \pi \ell/L$ with $L=512$. Namely, the plots at two smallest
$h$ values almost agree within the statistical error bars.

If we assume the corrections to scaling proposed by the standard theory, then
the asymptotic exponent $\lambda_{\parallel}$ can be quite well evaluated
by fitting these smallest--$h$ data as functions of $k$. Recall that these corrections are represented by
an expansion in powers of $k^{4-d}$ and $k^{d-2}$~\cite{Sch,Law1}, corresponding
to the expansion $\lambda_{\mathrm{eff}}(k) = \lambda_{\parallel} + \sum_{j \ge 1} a_j k^j$
in three dimensions. Accordingly, we have
plotted $\lambda_{\mathrm{eff}}(k)$ vs $k$ in Fig.~\ref{effexpo} and have 
fit these plots by a polynomial of $k$. The cubic fits well describe 
the shape of the theses curves within $k \in [k_{20},k_{70}]$ and yield
$\lambda_{\parallel} = 0.641(69)$ at $h=0.000875$ (dot--dot--dashed curve), 
$\lambda_{\parallel} = 0.703(77)$ at $h=0.0004375$ (dashed curve) and
$\lambda_{\parallel} = 0.687(65)$ at $h=0.00021875$ (solid curve).
In the latter case, the quadratic fit within $k \in [k_{20},k_{43}]$ (dot--dashed curve)
also is very good and looks plausible. It yields  $\lambda_{\parallel} = 0.694(63)$.
In fact, all these fits at $h=0.0004375$ and $h=0.00021875$ give well consistent
results within the statistical error bars. Therefore, our final combined estimate is 
$\lambda_{\parallel} = 0.69 \pm 0.10$, where $0.69$ is the rounded average
value of the two fits at the smallest field $h=0.00021875$, and the error bars
are roughly estimated as $\pm 0.10$ to include the statistical error,
as well as the systematical error due to small finite--$h$ effects.
This systematical error is assumed to be smaller than the discrepancy
between the estimates at $h=0.000875$ and $h=0.00021875$.
Our estimate $\lambda_{\parallel} = 0.69 \pm 0.10$ is clearly
inconsistent with the expected standard--theoretical value $\lambda_{\parallel}=1$.
In fact, the curves in Fig.~\ref{effexpo} deviate away from $1$, and
it looks very unlikely that effective exponent could converge to this value 
as an analytic function of $k$.

One has to note that corrections to scaling, proposed by the GFD theory~\cite{K2010},
are represented by an expansion in powers of $k^{2-\lambda_{\perp}}$, $k^{\lambda_{\perp} - \lambda_{\parallel}}$
and $k^{\lambda_{\parallel}}$. The existence of a small correction--to--scaling exponent $2 - \lambda_{\perp}$ 
is very important. It makes the extrapolation of the
$\lambda_{\mathrm{eff}}(k)$ plots unreliable, so that 
the true asymptotic value could be remarkably different from $\lambda_{\parallel} = 0.69 \pm 0.10$.
Nevertheless, the actual estimation is well justified as a test of consistency of
the standard theory.

We have analyzed also the effective transverse exponent. This, however, does not give
a better numerical evidence than those already considered in~\cite{KMR08,KMR10}.
We note only that the effective exponent (in the limit 
$\lim\limits_{h \to 0} \lim\limits_{L \to \infty}$), evaluated approximately from 
fits of the $G_{\perp}({\bf k})$ data within $[k,4k]$, apparently, has 
a maximum around $k \approx k_7$ with the maximum value 
$\lambda_{\mathrm{max}} \approx 1.976$. Thus,
$\lambda_{\perp} < \lambda_{\mathrm{max}} < 2$ most likely holds, 
in agreement with~(\ref{eq:pred1}).

\section{Conclusions}

\begin{enumerate}
\item
Different theoretical predictions for the power--law singularities and
critical exponents of the 
$n$--vector model (or $\varphi^4$ model) have been considered in
Secs.~\ref{sec:crex}, \ref{sec:nonp} and~\ref{sec:Goldstone}.
In particular, it has been discussed how to relate 
the possible values of the critical exponents,
proposed by the GFD (grouping of Feynman diagrams)
theory in~\cite{K_Ann01}, to specific $n$--vector models.
This approach has been considered as an alternative method
to the perturbative RG treatment.
\item
A non--perturbative proof concerning corrections to scaling
in the two--point correlation function of the $\varphi^4$ model has been provided in Sec.~\ref{sec:nonp}, 
showing that corrections to scaling proposed by the GFD theory rather than those of 
the perturbative RG theory can be correct.
\item
The known fits of the experimental specific
heat data very close to $\lambda$--transition point in liquid helium
have been critically reconsidered and tested in Sec.~\ref{sec:he}.
It turns out that the overall fits with the RG correction--to--scaling
exponent $\theta \simeq 0.529$ fail to give satisfactory results in certain
test of validity.
We have demonstrated also that these experimental
data can be very well interpreted with the critical exponents of the GFD theory,
according to (\ref{eq:result}), (\ref{eq:conj})  and (\ref{eq:correx}).
\item
Goldstone mode singularities in the 
$O(n)$ models have been discussed in Sec.~\ref{sec:Goldstone},
showing that recent Monte Carlo estimates are in agreement with
the theoretical predictions of the GFD theory. 
A new Monte Carlo evidence has been provided, according to which
the statement of the (old) standard theory that 
$G_{\parallel}({\bf k}) \sim k^{-\lambda_{\parallel}}$ with $\lambda_{\parallel}=1$
holds at $h=+0$ and $k \to 0$ in three dimensions does not look plausible, 
if corrections to scaling are such as proposed by this theory. In this 
case our estimation yields $\lambda_{\parallel} = 0.69 \pm 0.10$.
\end{enumerate}

\section*{Acknowledgements}

The comparison between theory and experimental data for liquid helium 
has been discussed with Reinhard Mahnke (Rostock) and Hans Weber (Lule{\aa}).
The Monte Carlo simulations were made possible by the facilities of the
Shared Hierarchical Academic Research Computing Network
(SHARCNET:www.sharcnet.ca) and by
the DEISA Consortium (www.deisa.eu), funded through the EU FP7
project RI-222919, for support within the DEISA Extreme Computing Initiative.

\end{document}